\documentclass[aps,prl,twocolumn,showpacs,preprintnumbers,nofootinbib,amsmath,amssymb]{revtex4}

\usepackage{graphicx}
\usepackage{dcolumn}
\usepackage{bm}
\usepackage{amsmath}

\newcommand{\beq}{\begin{eqnarray}}
\newcommand{\eeq}{\end{eqnarray}}

\newcommand{\eg}{{\it e.g.}}
\newcommand{\real}{{\sf I}\kern-.12em{\sf R}}
\newcommand{\comp}{{\sf I}\kern-.50em{\sf C}}
\newcommand{\unity}{{\sf I}\kern-.54em{\sf 1}}

\newcommand{\tr}{\mbox{Tr}}

\newcommand{\mbf}[1]{\ensuremath{\boldsymbol{#1}}}

\def\spose#1{\hbox to 0pt{#1\hss}}
\def\ltapprox{\mathrel{\spose{\lower 3pt\hbox{$\mathchar"218$}}
 \raise 2.0pt\hbox{$\mathchar"13C$}}}

\begin{document}

\title{Magnetic Susceptibility of Strongly Interacting Matter across the Deconfinement Transition}
\author{Claudio Bonati}
\email{bonati@df.unipi.it}
\affiliation{
Dipartimento di Fisica dell'Universit\`a
di Pisa and INFN - Sezione di Pisa,\\ Largo Pontecorvo 3, I-56127 Pisa, Italy}

\author{Massimo D'Elia}
\email{delia@df.unipi.it}
\affiliation{
Dipartimento di Fisica dell'Universit\`a
di Pisa and INFN - Sezione di Pisa,\\ Largo Pontecorvo 3, I-56127 Pisa, Italy}

\author{Marco Mariti}
\email{mariti@df.unipi.it}
\affiliation{
Dipartimento di Fisica dell'Universit\`a
di Pisa and INFN - Sezione di Pisa,\\ Largo Pontecorvo 3, I-56127 Pisa, Italy}

\author{Francesco Negro}
\email{fnegro@ge.infn.it}
\affiliation{Dipartimento di Fisica dell'Universit\`a
di Genova and INFN - Sezione di Genova,\\
 Via Dodecaneso 33, I-16146 Genova, Italy}

\author{Francesco Sanfilippo}
\email{francesco.sanfilippo@th.u-psud.fr}
\affiliation{Laboratoire de Physique Th\'eorique (Bat. 210) Universit\'e Paris
SUD, F-91405 Orsay-Cedex, France 
}
\date{\today}

\begin{abstract}
We propose a method to determine the total magnetic susceptibility of 
strongly interacting matter by lattice QCD simulations, and present
first numerical results for the theory with two light flavors, which 
suggest a weak magnetic activity in the 
confined phase and the emergence of strong paramagnetism in 
the deconfined, Quark-Gluon Plasma phase. 
\end{abstract}

\maketitle

{\it Introduction} -- Understanding the properties of strong interactions in the presence
of strong magnetic backgrounds is a problem of the utmost phenomenological
importance.
The physics of 
compact astrophysical
objects, like magnetars~\cite{magnetars}, 
of non-central heavy ion collisions~\cite{hi1,hi2,hi3,hi4} 
and of the early Universe~\cite{vacha,grarub}, 
involvs fields going from $10^{10}$ Tesla 
up to $10^{15-16}$ Tesla ($|e| B \sim 1$ GeV$^2$).
The problem is also relevant to a better comprehension of the 
non-perturbative properties of QCD and of the Stardard Model in general.
That justifies the recent theoretical efforts on the 
subject (see, e.g., Ref.~\cite{lecnotmag}).

Any material is characterized by the way it reacts
to electromagnetic external sources.
 For strongly interacting matter, 
such as that present in the early Universe and in the core of
compact astrophysical objects, or that created in heavy ion collisions,
the same questions as for any other medium can be posed.
Does it react linearly to magnetic backgrounds,
at least for small fields, and is it
a paramagnet or a diamagnet? How
the magnetic susceptibility
$\chi$ changes as a function
of the temperature $T$ and/or chemical potentials?

Despite the clear-cut nature of such
questions, a definite answer is still missing. 
Strong interactions in external fields can be 
conveniently explored by lattice QCD simulations;
various investigations 
have focussed till now on 
partial aspects, like the 
magnetic properties of the spin component~\cite{buivid,reg1} and
of the QCD vacuum~\cite{reg2}.
Most technical difficulties
are related to the fact that in a lattice setup, which usually adopts
toroidal geometries, the magnetic background is quantized.

In the following we propose a new method to overcome such difficulties
and present a first investigation for QCD with 2 light flavors
in the standard rooted staggered formulation, performed
at various values of the lattice spacing $a$ and of the quark masses.
Results show that $\chi$ is
small (vanishing within present errors) in the confined
phase, while it steeply rises above the transition,
i.e. the Quark-Gluon Plasma is paramagnetic.

{\it The method} -- The magnetic properties of a homogeneous
medium at thermal equilibrium
can be inferred from the change of its free energy
density, $f = F/V$, in terms
of an applied constant and uniform field:
\begin{equation}
\Delta f(B,T) = 
- \frac{T}{V} \log \left( \frac{Z(B,T,V)}{Z(0,T,V)} \right)
\label{freediff}
\end{equation}
where $Z = \exp(-F/T)$ 
is the partition function of the system, $B$ is 
the magnetic field modulus and $V$ is 
the spatial volume.
One usually deals directly with free energy derivatives, 
like the magnetization, 
which can be rewritten 
in terms of thermal expectation values and are extracted
more easily than free energy differences, whose
computation is notoriously difficult (see, e.g., Ref.~\cite{umbrella}).

However, in lattice simulations 
the best way to deal with a finite spatial volume, 
while minimizing finite size effects 
and keeping a homogeneous background field,
is to work on a compact manifold without boundaries, such as 
a 3D torus (cubic lattice with
periodic boundary conditions).
That leads to ambiguities in the presence 
of charged particles moving over the manifold, unless the total flux of the magnetic
field, across a section orthogonal to it, is quantized in units
of $2 \pi /q$, where $q$ is the elementary electric particle charge.
The same argument leads to Dirac quantization of the magnetic monopole charge, when considering a spherical surface
around it. In the case of the 3D torus, 
assuming $\mbf{B} = B\ \mbf{\hat z}$ and considering 
that for quarks $q = |e|/3$, 
one has~\cite{bound1,bound2,bound3,wiese}
\beq
|e| B = {6 \pi b}/{(l_x l_y)} 
\label{bquant}
\eeq
where $b$ is an integer and $l_x$, $l_y$ are the torus extensions 
in the $x, y$ directions.

Since $B$ is quantized, taking derivatives with respect to it is not 
well defined. New approaches can be found to get around 
the problem, like the anisotropy method~\cite{reg2}. However, one can 
still go back to Eq.~(\ref{freediff}) and consider finite free energy differences: this is our strategy, as explained in the following. Let us first recall
more details regarding the magnetic field on the lattice torus.

Electromagnetic fields
enter the QCD lagrangian through the covariant derivative of quarks,
$D_\mu = \partial_\mu + i\, g A^a_\mu T^a + i\, q A_\mu$, where 
$A_\mu$ is the electromagnetic gauge potential and $q$ 
is the quark electric charge. On the lattice, that corresponds
to adding proper $U(1)$ phases $u_\mu(n)$ to the $SU(3)$ parallel transports
entering the discretized Dirac operator,
$U_\mu(n) \to u_\mu(n) U_\mu(n)$, where $n$ is a lattice site.
A magnetic field $\mbf{B} = B\ \mbf{\hat z}$
can be realized, for instance, by a potential $A_y = B  x$
and $A_\mu = 0\ {\rm for}\ \mu\neq y$. In the presence of periodic 
boundary conditions, $B$ must be quantized as in Eq.~(\ref{bquant})
and proper b.c. must be chosen for fermions, to preserve gauge 
invariance~\cite{wiese}. The corresponding  $U(1)$ links are
\begin{equation}
\begin{aligned}
u_y^{(q)}(n) &= e^{i\, a^2 q B\, n_x} &=&\ e^{i\, 2 \pi b\,  n_x / (L_x L_y)} \\
u_x^{(q)}(n)|_{n_x = L_x} &= e^{-i\, a^2 q L_x B\, n_y} &=&\ e^{-i\, 2 \pi b\, n_y / L_y} 
\label{u1field}
\end{aligned}
\end{equation}
and $u_{\nu}(n)=1$ otherwise, where $n_{\mu}\in\{1,\ldots, L_{\mu}\}$, $L_\mu$ 
being the lattice extension along $\mu$; 
$b$ gets a factor -2 for
$u$ quarks with respect to $d$ quarks.

With this choice, a constant magnetic 
flux $a^2 B$ goes through all plaquettes in the $x y$ plane, apart from a ``singular'' plaquette located at 
$n_x = L_x$ and $n_y = L_y$, which is pierced by a flux $(1 - L_x L_y) a^2 B$, leading to a vanishing total flux through 
the $xy$ torus, as expected for a closed surface. In the continuum limit, 
that corresponds to a uniform 
magnetic field plus a Dirac string piercing the torus in one point:
like for Dirac monopoles, the string
carries all the flux away.   
However, if $B$ is quantized as in Eq.~(\ref{bquant}),
the string becomes invisible to all particles carrying electric charges 
multiple of $q$, and the phase of the singular plaquette becomes equivalent, 
modulo $2 \pi$, to that of all other plaquettes, i.e. the 
field is uniform.

Consider now the problem of computing finite free energy differences, $f(B_2) - f(B_1) = f(b_2) - f(b_1)$ 
where $b_1$ and $b_2$ are integers.
Several methods are known
to determine such differences in an efficient 
way (see, e.g., Ref.~\cite{thooft}): 
the general idea is to divide them into a sum of smaller, easily computable differences. 
We will consider infinitesimal differences and rewrite
\beq
f(b_2) - f(b_1) = \int_{b_1}^{b_2} \frac{\partial f(b)}{\partial b} \mathrm{d} b\, ,
\label{intfb}
\eeq
the idea being to determine the integral after computing the integrand on a grid of points, 
fine enough to keep systematic errors under control.

Let us clarify the meaning of $\partial f / \partial b$.
Generic real values of
$b$ correspond to a uniform field 
plus a visible Dirac string:
while this is not the physical situation we are interested in, 
it still represents a legitimate theory,  
interpolating between integer values of 
$b$. 
In practice, 
we are extending a function, originally defined on integers, 
to the real axis, and
then we are integrating its derivative,
$\partial f / \partial b$,
between integer values
to recover the original function: 
as long as the extension is analytic, as
always possible on a finite lattice, the operation is well defined
(see the Appendix for an explicit check).

Therefore, ${\partial f}/{\partial b}$ has no direct relation
with the magnetization, even for integer $b$.
For the particular interpolation adopted, 
a large contribution to it
comes from the string itself, leading to a characteristic
oscillating behavior; since $f(b)$
has a local minimum when the string becomes invisible,
${\partial f}/{\partial b}$ vanishes for integer $b$.

{\it Renormalization} --  The procedure described above gives access to 
$\Delta f(B,T)$, defined in Eq.~(\ref{freediff}), however
we have to take care of divergent contributions.
Indeed, $B$-dependent divergences do not cancel
when taking the difference $\Delta f$, and must
be properly subtracted, with possible ambiguities 
related to the definition of the vacuum energy in the presence of a 
magnetic field. For $T = 0$, the prescription of Ref.~\cite{reg2} is to 
subtract all terms quadratic in $B$, so that, by definition, 
the magnetic properties of the QCD vacuum are of higher order in $B$.

In the following, we are not interested 
in the magnetic properties of vacuum, but only in those of 
the strongly interacting thermal medium, 
which may be probed experimentally.
Therefore, our prescription is to compute the following quantity:
\beq
\Delta f_R (B,T) = 
\Delta f(B,T) - \Delta f(B,0) 
\label{subtr}
\eeq
which is properly renormalized, since all vacuum (zero $T$) contributions have been subtracted
and no further divergences, depending both on $B$ and on $T$, appear
(see, \eg, the discussion in Refs.~\cite{reg0, reg2}). 
Clearly, divergences are really removed only if 
the contributions to Eq.~\eqref{subtr} 
are evaluated at a fixed value of the lattice spacing.
The small field behavior of $\Delta f_R$ 
will give access to 
the magnetic susceptibility of the medium.

{\it Effects of QED quenching} -- 
For small fields and for a linear, homogeneous and isotropic 
medium, the magnetization 
$\mbf{\mathcal{M}}$ is proportional to the total 
field $\mbf{B}$ acting on the medium, 
$\mbf{\mathcal{M}} = \tilde \chi \mbf{B}/ \mu_0$
(using SI units), where 
$\tilde \chi$ is the susceptibility. The 
relation can also be expressed as $\mbf{\mathcal{M}}=\chi\mbf{H}$,
where $\mbf{H} = \mbf{B}/\mu_0 - \mbf{\mathcal{M}}$ and
the relation $\chi = \tilde \chi / (1 - \tilde \chi)$ holds
between the two different definitions of susceptibility.

The change in the free energy density is usually written in the form 
$\Delta f =\int \mbf{H}\cdot\mathrm{d}\mbf{B}$
(see, \eg, Ref.~\cite{LL} \S 31).
However, in $\Delta f_R$ 
the energy of the magnetic field alone is subtracted, hence 
the proper expression is:
$\Delta f_R = -\int \mbf{\mathcal{M}}\cdot\mathrm{d}\mbf{B}$.
Taking into account 
$\mbf{\mathcal{M}} = \tilde\chi \mbf{B}/\mu_0$ we get,
in the limit of small fields,
\begin{equation}
\Delta f_R =-\frac{\tilde\chi}{\mu_0}\int \mbf{B}\cdot\mathrm{d}\mbf{B}=-\frac{\tilde\chi}{2\mu_0}\mbf{B}^2\  \equiv -\frac{\hat\chi}{2} (e \mbf{B})^2\
\label{intfree2} \, .
\end{equation}
The field $\mbf{B}$ in the last equation
is the total field felt by the particles of the medium,
i.e. that entering the Dirac matrix:
since in our setup the dynamics of 
electromagnetic fields is quenched,
it coincides with the
external field added to the system,
i.e. we do not have to add the field generated by the
magnetization itself.
Last quantity introduced in Eq.~(\ref{intfree2}),
$\hat\chi$, will be used for $\Delta f_R$ and $(e B)^2$ both measured in 
natural units.  

\begin{table}[t!]
\begin{center}
\begin{tabular}{|c|c|c|c|c|c|c|c|c|}
\hline
$L_s$ & $L_t$ & $\beta$ & $am$ & $a$[fm] & $m_{\pi}$ & $T$ & $\tilde{\chi}\times 10^3$\rule{0mm}{3mm} & $\hat\chi \times 10^2$\rule{0mm}{3mm} \\
\hline
\hline
20 & 4 & 5.4075 & 0.00334 & 0.188 & 195 & 262 & 1.89(21) & 2.06(23) \\ \hline
16 & 4 & 5.4342 & 0.00584 & 0.17 & 275 & 290 & 2.04(13) & 2.22(15)  \\ \hline
16 & 6 & 5.4342 & 0.00584 & 0.17 & 275 & 193 & 0.70(15) & 0.76(16) \\ \hline
16 & 8 & 5.4342 & 0.00584 & 0.17 & 275 & 145 & 0.23(23) & 0.25(25) \\ \hline
24 & 4 & 5.527 & 0.0146 & 0.141 & 480 & 349 & 2.69(20)  & 2.93(22) \\ \hline
24 & 6 & 5.527 & 0.0146 & 0.141 & 480 & 233 & 1.42(16)  & 1.55(18) \\ \hline
24 & 8 & 5.527 & 0.0146 & 0.141 & 480 & 175 & 0.49(21)  & 0.53(22) \\ \hline
24 & 10 & 5.527 & 0.0146 & 0.141 & 480 & 140 & 0.15(20) & 0.16(22) \\ \hline
16 & 4 & 5.453 & 0.02627 & 0.188 & 480 & 262 & 1.54(10) & 1.68(10) \\ \hline
16 & 6 & 5.453 & 0.02627 & 0.188 & 480 & 175 & 0.21(11) & 0.23(12) \\ \hline
16 & 8 & 5.453 & 0.02627 & 0.188 & 480 & 131 & 0.05(11) & 0.05(12) \\ \hline
16 & 4 & 5.3945 & 0.0495 & 0.24 & 480 & 205 & 0.51(7)   & 0.56(8) \\ \hline
16 & 8 & 5.3945 & 0.0495 & 0.24 & 480 & 103 & 0.00(8)   & 0.00(9)\\ \hline
\end{tabular}
\end{center}
\caption{Lattice parameters and results. $T$ and 
$m_\pi$ are in MeV units, $\tilde\chi$ in SI units, 
while $\hat\chi$ for both $\Delta f_R$ and
$(eB)^2$ measured in natural units (see Eq.~(\ref{intfree2})). 
}
\label{tab1}
\end{table}

{\it Numerical results} --
As a first application of our method, we consider
$N_f = 2$ QCD with fermions in
the standard rooted staggered formulation, with
each quark described
by the fourth root of the fermion determinant. The 
partition function reads:
\beq
Z \equiv \int \mathcal{D}U e^{-S_{G}} 
\det D^{1\over 4} [U,q_u]
\det D^{1\over 4} [U,q_d]
\:
\label{partfun1}
\eeq
\begin{eqnarray}
D^{(q)}_{i,j} \equiv a m \delta_{i,j} 
&+& {1 \over 2} \sum_{\nu=1}^{4} \eta_\nu(i) \left(
u_\nu^{(q)}(i)\ U_{\nu}(i) \delta_{i,j-\hat\nu}
\right. \nonumber \\ &-& \left.
u^{*(q)}_\nu{(i - \hat\nu)}\ U^{\dag}_\nu{(i-\hat\nu)} \delta_{i,j+\hat\nu} 
\right) \:
\label{fmatrix1}
\end{eqnarray}
$\mathcal{D}U$ is the integration over $SU(3)$ gauge link
variables, $S_G$ is the plaquette action,  
$i$, $j$ are lattice site indexes,
$\eta_{\nu}(i)$ are the staggered
phases. The quark charges are
$q_u = 2 |e|/3$ and $q_d = - |e|/3$. 
The density of the integrand in Eq.~(\ref{intfb}) can be expressed as
\begin{equation}
M \equiv a^4 \frac{\partial f}{\partial b}=\frac{1}{4 L_t L_s^3}
 \sum_{q = u,d} \Big\langle \tr\Big\{ 
\frac{\partial D^{(q)}}{\partial b} {D^{(q)}}^{-1} \Big\} \Big\rangle\, 
\label{M_def}
\end{equation}
where $L_s$ and $L_t$ are the temporal and spatial sizes
($T = 1/(L_t a)$).
We stress again that
$M$ is just 
the derivative of the free energy interpolation
and has no direct physical interpretation.

\begin{figure}[t!]
\includegraphics[width=0.9\columnwidth, clip]{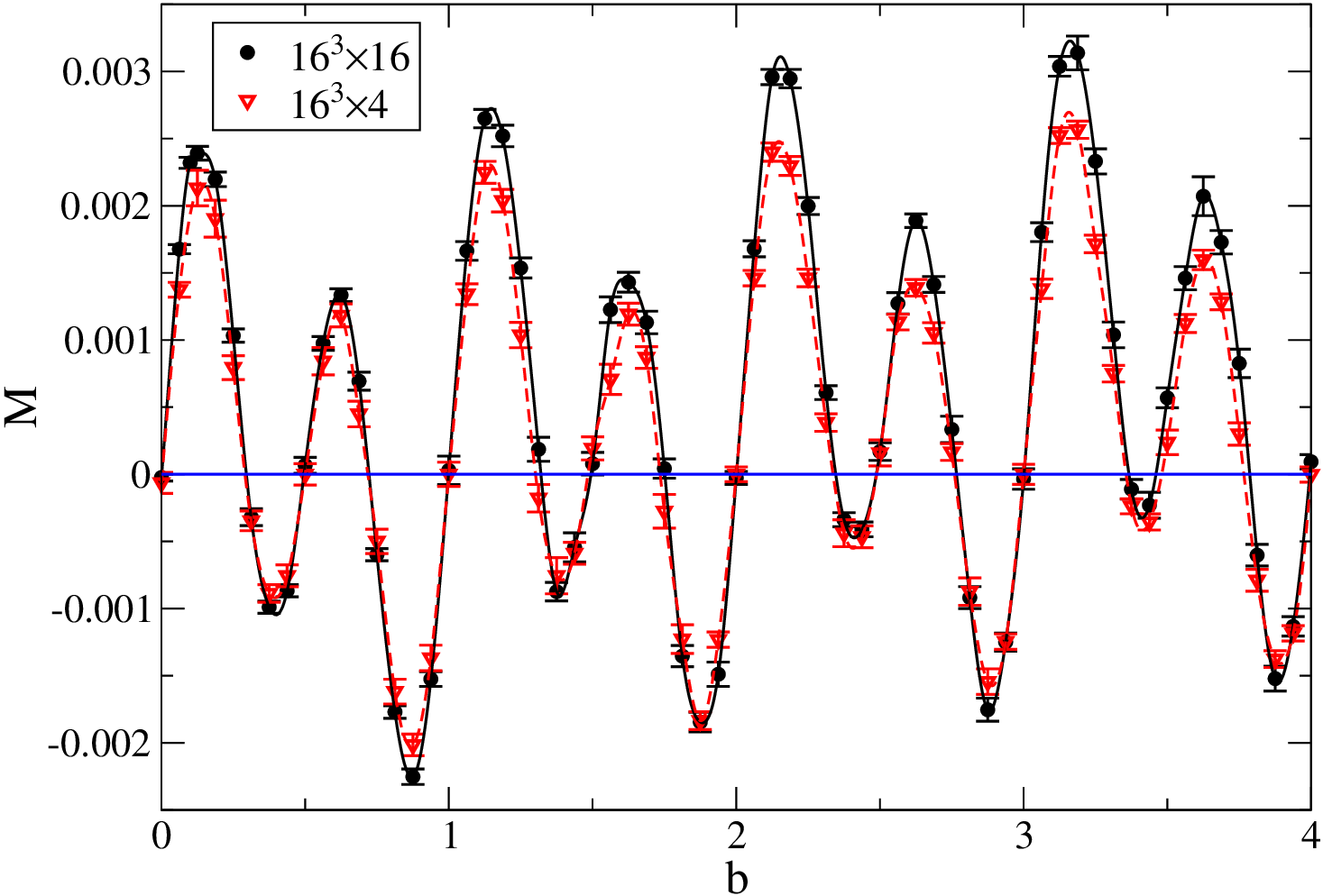}
\caption{$M$ computed on $16^4$ and $16^3 \times 4$ lattices, 
with $a\approx 0.188$\,fm and $m_{\pi}\approx 480$\,MeV.
The lines are third order spline interpolations.}
\label{M_fig}
\end{figure}

We have explored different lattice spacings and pseudo-Goldstone
pion masses, by tuning 
the inverse gauge coupling $\beta$ and $a m$ according 
to Ref.~\cite{blum} (the magnetic background does not modify 
$a$~\cite{reg0, thetaeff}), and different values 
of $L_s$ and $L_t$ (see Table~\ref{tab1}).
For the explored sets, the pseudocritical temperature
$T_c$ is in the range 160-170 MeV~\cite{demusa}.
We have adopted a 
Rational Hybrid Monte-Carlo (RHMC)
algorithm implemented on GPU cards~\cite{gpu}, with statistics 
of $O(10^3)$ molecular dynamics (MD) time units
for each $b$. $M$
has been measured every 5 trajectories, of one MD time unit each, 
adopting a noisy estimator, with $10$ random vectors for each measure.

Fig.~\ref{M_fig} shows an example of the determination of $M$,
for the first 4 quanta of $B$, for one parameter set
and for
$L_t = 4$ and  $16$, the latter being taken 
as our $T \sim 0$ 
reference value. 
Oscillations between successive quanta can be related 
to the presence of the string: the two visible harmonics are associable
with the $d$ and $u$ quark contributions, which feel the string differently.

Despite the unphysical oscillations, $M$ is smooth enough 
to perform a numerical integration:
that is done by using a spline interpolation over 
16 equally spaced determinations of $M$ 
for each  quantum;
errors are estimated by means of a bootstrap analysis.
We checked that variations due to 
different integration schemes, or to
different interpolating strategies and densities, 
always stay well within the estimated errors, 
so that the integration procedure is very robust
(see the Appendix for details).

\begin{figure}[t!]
\includegraphics[width=0.9\columnwidth, clip]{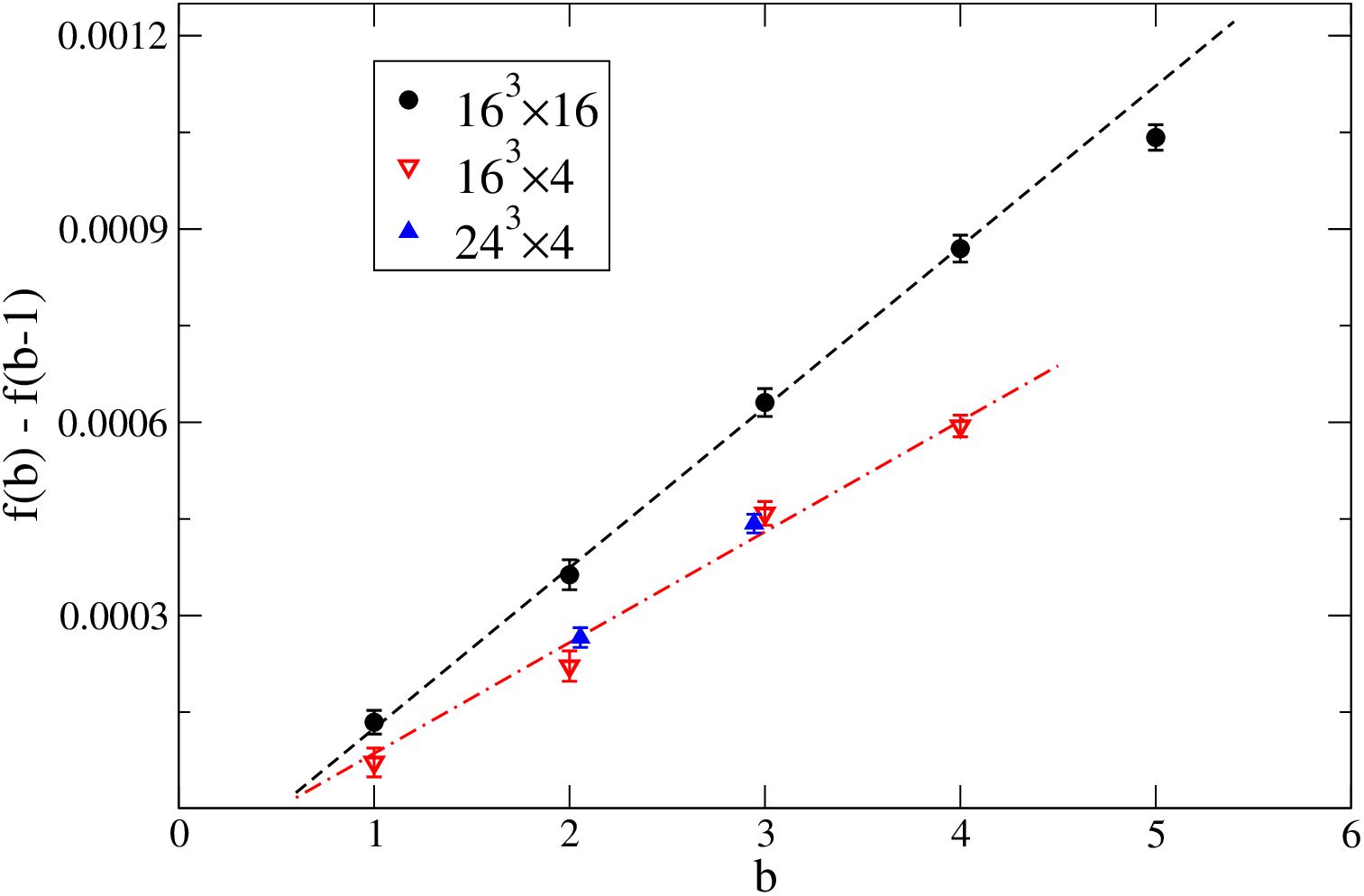}
\caption{$f(b) - f(b-1)$ computed from data in 
Fig.~\ref{M_fig}, together with best fits obtained, for $b \leq 4$,
according to $c_2\, (2b - 1)$ (see Eq.~\eqref{dDF_fit}). 
Two further, properly rescaled data points are reported
from a $24^3 \times 4$ lattice.}
\label{F_fig}
\end{figure}

To obtain the $\mathcal{O}(B^2)$ term in 
$\Delta f_R (B,T)$, we have 
determined the $\mathcal{O}(B^2)$ contributions to both $\Delta f(B,T)$ 
and $\Delta f(B,0)$, then we have subtracted them; 
consistent results are obtained if the subtraction is performed first.
Assuming that 
$a^4 \Delta f(b) \equiv c_2\, b^2 + O(b^4)$ holds for integer $b$, 
$c_2$ is conveniently determined by looking at the 
differences between successive quanta,
\begin{equation}
a^4\, (f(b) - f(b-1)) \equiv \int_{b-1}^b M(\tilde b) \mathrm{d}\tilde{b}\, 
\simeq\, c_2\, (2b-1)\, , 
\label{dDF_fit}
\end{equation}
so that the whole difference $f(b) - f(0)$ is not needed, and 
fitted data have independent errors, since
the integration uncertainties do not propagate 
between consecutive quanta.

The finite differences obtained from the data in Fig.~\ref{M_fig}
are reported in Fig.~\ref{F_fig}. A fit to
$c_2\, (2b - 1)$ works well for $b \leq 4$, yielding
$c_2 = 0.861(20) \times 10^{-4}$  ($\chi^2/{\rm d.o.f.} = 5.4/3$)
for $L_t = 4$ and 
$c_2 = 1.309(21) \times 10^{-4}$  ($\chi^2/{\rm d.o.f.} = 0.5/3$)
for $L_t = 16$. A fit in the same range to a generic power law 
$f(b) \propto b^\gamma$ returns, e.g. for $L_t = 16$, $\gamma = 1.99(3)$, 
excluding behaviors different from a linear response medium
(e.g., ferromagnetic-like).
Two further data points are reported from 
a $24^3 \times 4$ lattice, after proper rescaling, to check 
for spatial volume independence. 
Finally, we get $a^4 \Delta f_R = {c_2}_R b^2 + O(b^4)$, with
${c_2}_R = - 0.448(29) \times 10^{-4}$.

\begin{figure}[t!]
\includegraphics[width=0.9\columnwidth, clip]{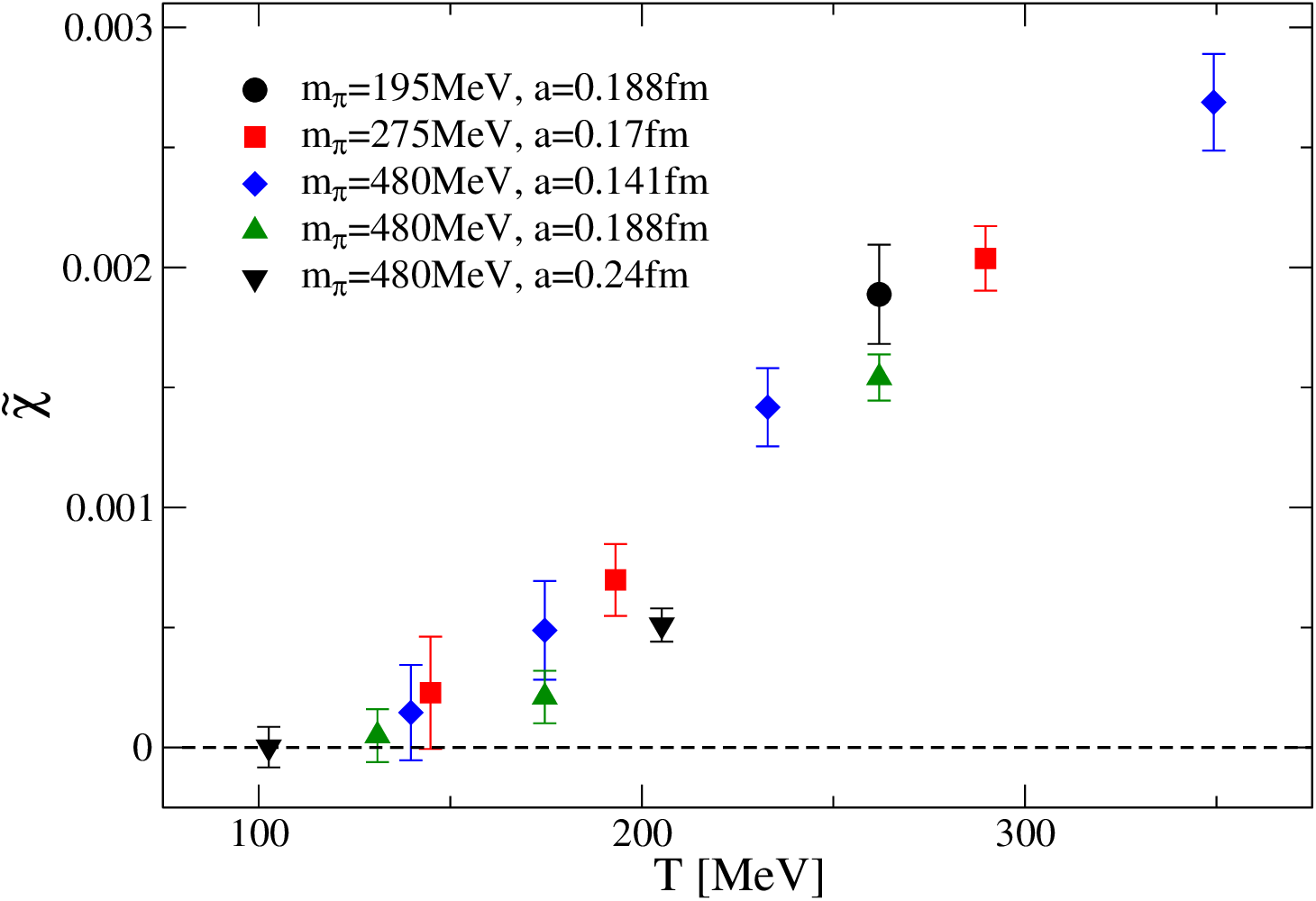}
\caption{Susceptibility 
(SI units) as a function of $T$, for different values of $m_\pi$ and $a$.
}
\label{final_fig}
\end{figure}

The determination of $\tilde \chi$ from Eq.~(\ref{intfree2})
requires a conversion into physical units for $\Delta f_R$ 
and $b$, according to Eq.~(\ref{bquant}). The result is 
\beq
\tilde \chi = - \frac{|e|^2 \mu_0 c}{18 \hbar \pi^2}\, L_s^4\, {c_2}_R \, ,
\eeq
in SI units ($\hbar$ and $c$ have been reintroduced explicitly). 
We obtain
$\tilde \chi = 0.00154(10)$, which indicates strong paramagnetism
when compared with those of ordinary materials~\cite{endnote1}.
Instead, adopting natural units,
one obtains
$\hat\chi = - L_s^4\, {c_2}_R\, / (18 \pi^2) = 0.0168(10)$
 (see Eq.~(\ref{intfree2})).
The same procedure described  in detail above has been repeated for 
all combinations of $m_\pi$, $T$ and $a$ reported in  
Table~\ref{tab1}. Results are shown in Table~\ref{tab1}
and Fig.~\ref{final_fig}.

{\it Discussion} --
Fig.~\ref{final_fig} shows that $\tilde\chi$ is compatible with
zero, within errors, in the confined phase, while it rises roughly linearly
with $T$ in the deconfined one. 
The drastic increase of $\tilde\chi$, which is naturally associable 
to quark liberation, implies a proportional increase of 
the $B$-dependent (quadratic) contribution to the pressure.
Such results are confirmed by a recent approach based on 
a Taylor expansion in $B$~\cite{levkova}.

At $m_\pi \simeq 480$ MeV, we performed a continuum extrapolation 
according to:
$\tilde\chi = A\, (T-\tilde T) + A'\, a^2$, 
which effectively describes all data with
$T > 170$ MeV ($\chi^2/{\rm d.o.f} = 2.5/3$),
with coefficients 
$A=1.37(12) \times 10^{-5}$ MeV$^{-1}$, $A'= -3.80(15) \times 10^{-4}$ GeV$^2$
and $\tilde T = 126(16)$ MeV 
(multiplication of $A$ and $A'$ by 10.9 provides the conversion to
$\hat\chi$).
When $m_\pi$ decreases, a modest increase of $\tilde\chi$ 
is observed; one might expect a further slight increase
after continuum extrapolation also in this case.

The computation proposed and first performed in this study
surely claims for an extension to the physical case. 
Our results do not suggest drastic changes
when decreasing $m_\pi$. The inclusion of the 
strange quark, instead, may increase
$\tilde\chi$ by about 20\%. 
Indeed,
separating the contributions 
to $\tilde\chi$ from $u$ and $d$ quarks, see Eq.~(\ref{M_def}),
one obtains $\tilde\chi_u \sim 4\, \tilde\chi_d$, as expected naively
on a charge counting basis
(see the Appendix for details), and one may expect
$\tilde\chi_s \sim \tilde\chi_d$.

Future studies should also clarify the behavior
of $\tilde\chi$ around $T_c$ and its relation
to confinement/deconfinement: while present results
are compatible with zero in the confined phase, 
improved determinations could better fix the magnitude
and sign of $\tilde\chi$ below $T_c$.
An extension to the case of chromomagnetic 
fields may be interesting as well~\cite{bari}.

Finally, we notice, following Ref.~\cite{fraga}, that 
the strong paramagnetic behavior,
rising with $T$, in the deconfined phase, and the fact that
finite $a$ effects tend to diminish it, may explain
the lowering of the pseudocritical temperature with $B$~\cite{reg0},
and why a different behavior was observed on coarse 
lattices~\cite{demusa, ilgen}.

\noindent {\bf Acknowledgements:}
We thank E.~D'Emilio, E.~Fraga and S.~Mukherjee
for useful discussions. 
Numerical computations have been performed on computer facilities
provided by INFN, in particular on two GPU farms in Pisa and Genoa
and on the QUONG GPU cluster in Rome.

\appendix

\section{APPENDIX}
\label{appendA}

In the following 
we will discuss a few additional results from our
simulations, in order to better elucidate some details of our 
procedure and to check for possible systematic effects.

The first question one could ask regards the stability of the results
against a change of the integration procedure, adopted to exploit
Eq.~(\ref{intfb}).
To that purpose, we report in Table~\ref{tab_inter} the results
of the integration over one given quantum of field (reference parameters are 
the same as for Fig.~\ref{M_fig}), 
obtained by varying the order of the spline interpolation used by
the integrator and/or the number of points over which $M$ is evaluated. 
It turns out that 
the integration is extremely stable, with variations well below
statistical fluctuations.

\begin{table}[h]
\begin{center}
\begin{tabular}{|c|c|c|}
\hline
$s$ & $16$ points & $32$ points \\
\hline
\hline
1  &  0.000596(16)  &  0.000594(12) \\
\hline
2  &  0.000594(17)  &  0.000593(12) \\
\hline
3  &  0.000592(17)  &  0.000594(12) \\
\hline
4  &  0.000592(17)  &  0.000594(13) \\
\hline
\end{tabular}
\end{center}
\caption{Result of the integration of $M$ between $b=3$ and $b=4$ on a $16^3\times 4$ lattice
(with $m_{\pi}\approx 480$\,MeV and $a\approx0.188$\,fm) using different methods:
$s$ is the degree of the spline interpolation and the integral is computed 
starting from meshes of
$16$ or $32$ equally spaced points.}
\label{tab_inter}
\end{table}

\begin{figure}[t!]
\includegraphics[width=0.9\columnwidth, clip]{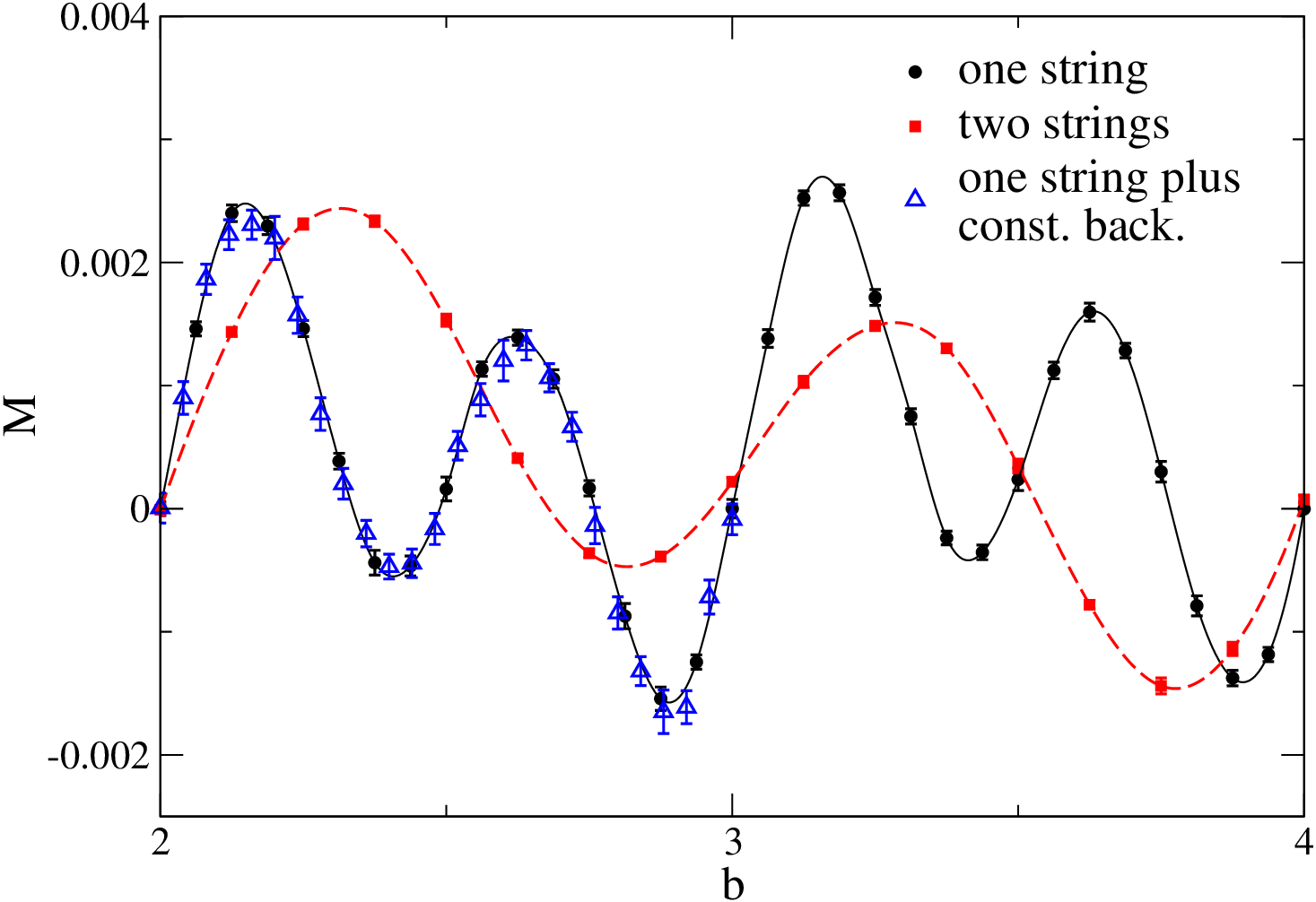}
\caption{$M$ computed between $b = 2$ and $b = 4$, and for the same
lattice parameters as in Table~\ref{tab_inter}, for three different
interpolations of the free energy (see text).}
\label{twostring1}
\end{figure}

\begin{figure}[t!]
\includegraphics[width=0.9\columnwidth, clip]{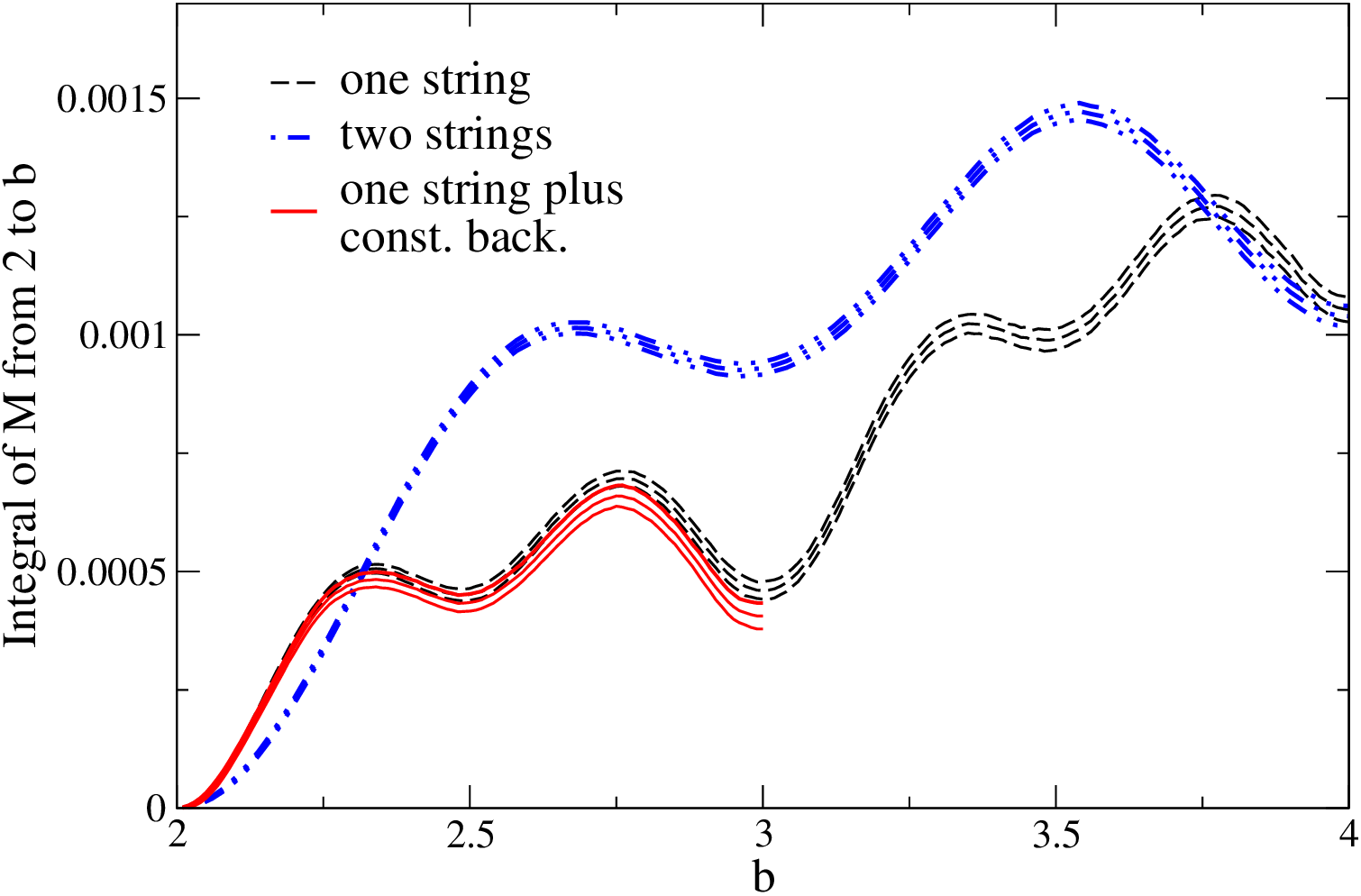}
\caption{Cumulative integrals of the three functions reported in 
Fig.~\ref{twostring1}.}
\label{twostring2}
\end{figure}

A different issue regards the stability of the result 
against a variation of the 
free energy interpolation. The simplest, alternative 
interpolation, consists in allowing for two (or more) different
Dirac strings at the same time, located
in different points. That is achieved by 
superposing two 
$U(1)$ fields like that in Eq.~(\ref{u1field}),
but with one of them shifted in one or two coordinates,
so as to move the location of the string: in this way 
one obtains an interpolation between two consecutive,
even quanta, however odd quanta are not possible any more. 
In Fig.~\ref{twostring1} we show, as 
an example, the values 
of $M$ between $b = 2$ and $b = 4$ obtained for the standard
and for the alternative interpolation described above: in the 
latter case, the two Dirac strings pierce the $x,y$ plaquettes 
located at  $(n_x,n_y) = (L_x,L_y)$ and $(L_x,L_y/2)$, respectively.
The corresponding cumulative integrals are reported in 
Fig.~\ref{twostring2}: they coincide, within errors, for 
values of $b$ where strings become invisible for both interpolations,
proving the stability of the procedure.

As a further, alternative interpolation, we have also tried
to modify the standard one by adding
a uniform  $U(1)$ background which disappears for integer
values of $b$. Results are shown in Figs.~\ref{twostring1}
and \ref{twostring2} as well, for one single 
quantum and for the case 
where a phase $\exp(i 2 \pi b)$ is added to all links along the 
$y$ direction: they are perfectly compatible with those
from the standard interpolation, even if a statistics larger 
by a factor 10
had to be used, due to the fact that the observable is much noisier 
in this case.

\begin{figure}[t!]
\includegraphics[width=0.9\columnwidth, clip]{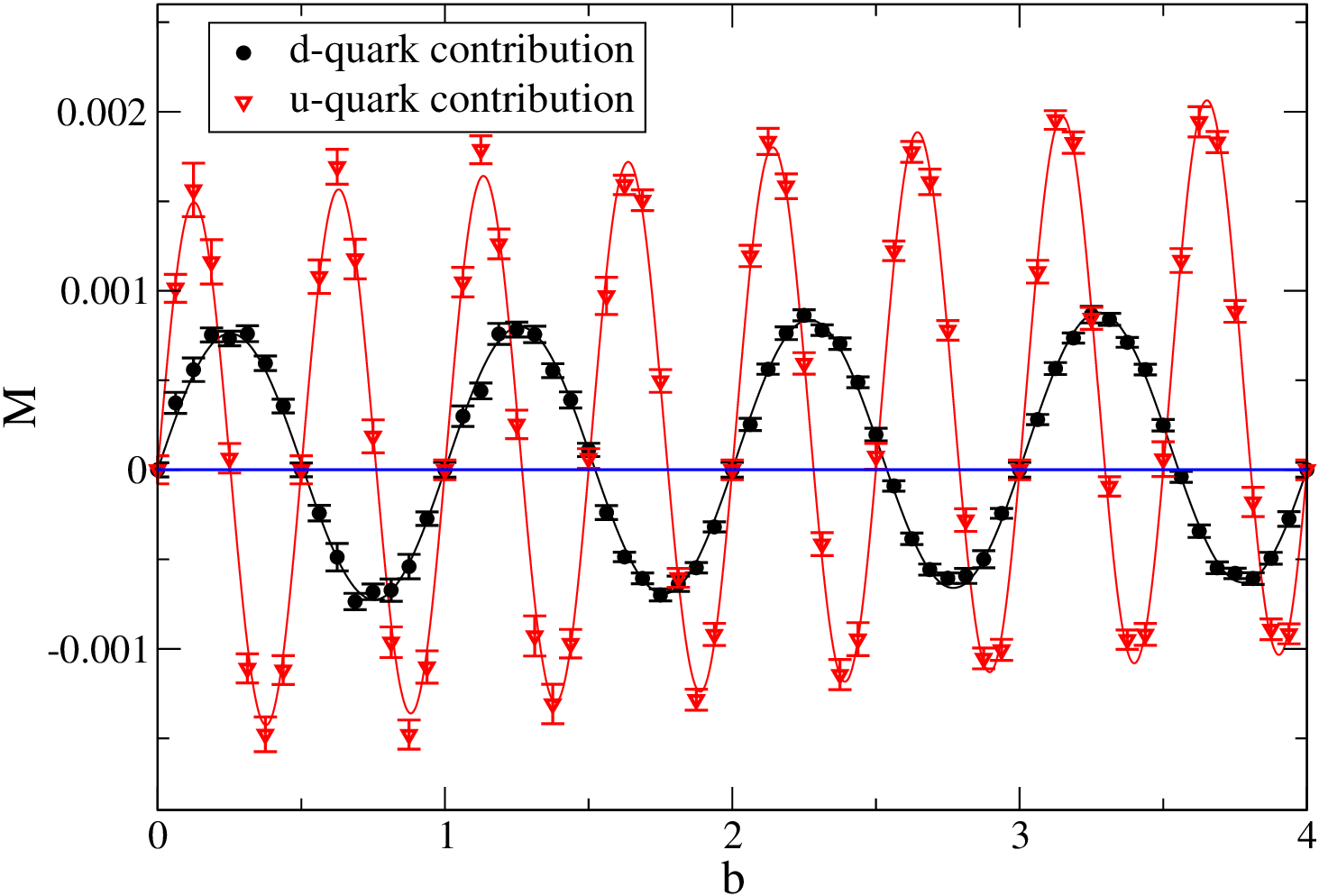}
\caption{Contributions to $M$ from the $u$ and $d$
quarks, computed on a $16^3 \times 4$ lattice, 
with $a\approx 0.188$\,fm and $m_{\pi}\approx 480$\,MeV.
The continuous lines are best fits according to Eq.~(\ref{fitfun}).}
\label{udsep}
\end{figure}

Finally, since the observable $M$ is made up of two different
terms, $M_u$ and $M_d$, 
coming from each quark determinant (see Eq.~(\ref{M_def})),
it is interesting to see how the two contributions look like. That is 
shown in Fig.~\ref{udsep}, where the same data shown 
in Fig.~\ref{M_fig}
for the $16^3 \times 4$ lattice have been split accordingly.
$M_u$ and $M_d$ present very similar oscillations, apart from a 
factor two in the frequency, which can be trivially associated 
to the electric charge ratio of the two quarks. It is interesting
that results can be described by the simplest function which can 
be devised by requiring that: {\it i)} it vanishes at points where the 
string becomes invisible to the corresponding quark; {\it ii)} it has 
a non-vanishing integral between any consecutive pair of such points;
{\it iii)} it is an odd function of $b$, as required by the 
charge conjugation symmetry present at $b = 0$. Such function is 
\beq
M^{\rm try}_q = A\, \sin \left( 2 \pi \frac{q}{q_d} b \right) 
+ A'\, b\, \left( 1 - \cos\left(2 \pi \frac{q}{q_d} b \right) \right) \, ,
\label{fitfun}
\eeq
where $q = q_u$ or $q = q_d$, and fits very well all 
data in Fig.~\ref{udsep}, with 
$\chi^2/{\rm d.o.f.} = 0.81$ for the $d$ quark and 
$\chi^2/{\rm d.o.f.} = 1.10$ for the $u$ quark. 
It is easy to check that the integral of such function
between 0 and integer values of $b$ equals $A'\, b^2/2$
(fit values for $A'$ are compatible 
with those from the standard spline integrators), therefore
deviations from such simple description are expected as soon
as the corrections to the quadratic behavior of 
$\Delta f$ become visible.

Data obtained for $M_u$ and $M_d$ can be integrated separately 
for each lattice setup, in this way also the renormalized free energy
and the corresponding magnetic susceptibility
can be separated into  two different contributions, 
$\tilde\chi = \tilde\chi_u + \tilde\chi_d$.  For the case shown explicitly in 
Fig.~\ref{udsep}, one obtains $\tilde\chi_u = 0.00122(9)$ and
$\tilde\chi_d = 0.000315(30)$. 
Even 
if one cannot strictly speak of $u$ and $d$ contributions,
because of  quark loop effects
which mix the two terms, 
it is nice
to observe that $\tilde\chi_u/\tilde\chi_d = 3.9(4) \sim (q_u/q_d)^2$, in agreement with 
a naive charge counting rule. Similar results are obtained
for the other values of $T$ and $m_\pi$ explored in this study.


\begin{thebibliography}{99}
\bibitem{endnote1}
As an example, we report $\hat\chi \simeq 2.8 \times 10^{-4}$ for Platinum and
$\hat\chi \simeq 3.9 \times 10^{-3}$ for Liquid Oxygen.
\end{thebibliography}
\end{document}